\documentclass[conference,10pt]{IEEEtran}
\usepackage[dvips]{color}
\usepackage{epsf}
\usepackage{times}
\usepackage{epsfig}
\usepackage{graphicx}

\usepackage{amsmath}
\usepackage{amssymb}
\usepackage{amsxtra}
\usepackage{amsthm}
\usepackage{bbm}

\usepackage{here}
\usepackage{rawfonts}
\usepackage{times}
\usepackage{url}
\usepackage{cite}
\usepackage{comment}
\usepackage[utf8]{inputenc}
\usepackage{caption}
\usepackage{subcaption}

\usepackage{pstricks}
\usepackage{algorithm}
\usepackage{algpseudocode}
\usepackage{lipsum}
\usepackage{mathtools}

\usepackage{geometry}
\makeatletter 

\begin{document}
\title{\huge Semi-Supervised  Learning for Channel Charting-Aided IoT  Localization in Millimeter Wave Networks}  
\author{ 
	\IEEEauthorblockN{ Qianqian Zhang and Walid Saad }
	
	\IEEEauthorblockA{\small 
		Bradley Department of Electrical and Computer Engineering, Virginia Tech, VA, USA,
		Emails: \url{{qqz93,walids}@vt.edu}. 
		%\thanks{This research was supported by the U.S. National Science Foundation under Grant CNS-1526844.}
	}
} 
\maketitle

\setlength{\columnsep}{0.55cm}
\begin{abstract}
In this paper, a novel framework is proposed for channel charting (CC)-aided localization  in millimeter wave  networks.  
In particular, a convolutional autoencoder model is proposed to estimate the three-dimensional location of wireless  user equipment (UE), based on  multipath channel state information (CSI), received by different base stations.   
In order to learn the radio-geometry map and capture the relative position of each UE, an autoencoder-based channel chart is constructed in an unsupervised manner,  such that neighboring UEs in the physical space will  remain close in the channel chart.  
Next, the channel charting model is extended to a semi-supervised framework, where the autoencoder is divided into two components: an encoder and a decoder, and each component is  optimized individually, using the labeled CSI dataset with  associated location information, to further improve  positioning accuracy.   
Simulation results show that the proposed  CC-aided  semi-supervised localization yields a higher accuracy, compared with existing supervised positioning and conventional unsupervised CC approaches.

\end{abstract}

\IEEEpeerreviewmaketitle

\section{Introduction}

Next-generation wireless systems will inevitably rely on millimeter wave (mmWave) bands to meet the increasing  demand of wireless connectivity from end user equipment (UE) \cite{shahmansoori2017position}.  
Given their large available bandwidths and by using multiple-input-multiple-output (MIMO) technologies, wireless mmWave systems can support a variety of new applications, including ultra-high-speed low-latency communications and localization-based services. 
In order to facilitate reliable data demodulation, MIMO transmissions require accurate knowledge of the channel state information (CSI). 
The CSI  not only reflects the  multipath propagation of wireless links, but also captures possible reflection in wideband  channels. 
Thus, by monitoring the pattern shift of CSI features, it is possible to infer the position of the transmitting UE   \cite{xiao2013pilot}. 
Indeed, the CSI of mmWave  channels contains rich knowledge of the wireless communication environment, allowing  a mapping from the radio space to the physical geometry for UE positioning purposes. 

Compared with emerging CSI-based localization, conventional positioning approaches become inadequate for the new applications, such as Internet of things (IoT).
For example, in dense urban areas and indoor scenarios, the  signal from Global Positioning System (GPS) is subject to large errors.  
Meanwhile, due to power limitation, IoT devices cannot be equipped with a GPS module for any position signal. %However, the localization for these IoT   may be required for the purpose of near-field information collection, equipment update and recycling, and checking the operation status \cite{khelifi2019survey}.  
In order to allow a GPS-free localization in a cellular network, a trilateration positioning algorithm is developed, using  the received signal strength (RSS),   to estimate the radial distance. % %an alternative   scheme is to employ   
However,  RSS measurements are distorted, due to the multipath fading and the highly dynamic states of mmWave channels, thus introducing positioning errors in radial distance estimation. % Meanwhile, the selection of the best three base stations (BSs) with the lowest localization error is not been explored.  
To address the localization challenges for  emerging IoT applications,  a recent approach, called  fingerprinting localization, is proposed to measure the multipath parameters of the wireless communications as fingerprints, and then, infer the locations of wireless users \cite{zhou2017grassma}. %address these limitations and %distinguish
However, the construction of an RSS radio map  for location fingerprinting requires extensive channel measurements, and it cannot deal with dynamic communication environments.  %which is time-consuming and cost-inefficient. % effort, which is significantly labor-intensive and time-consuming. % requires significant channel measurements, and  fingerprinting-based localization relies on a multiple of densely-spaced BSs.  
%Thus, fingerprinting-based localization is limited in the small-area  communication systems.  

\subsection{Prior Works} 
In order to  enable flexible and cost-efficient positioning in  mmWave networks,  a number of CSI-based  approaches were developed in  \cite{zhou2017grassma, olivier2016lightweight,palacios2017jade, chen2017pseudo, studer2018channel, geng2020multipoint, deng2018multipoint, pan2011tracking,chidlovskii2019semi}.  
Based on the directional communication model of mmWave frequencies, geometry optimization-based localization was investigated using the  angle difference-of-arrival \cite{olivier2016lightweight}, angle-of-arrival (AoA) \cite{palacios2017jade}, and time-of-arrival (ToA) \cite{chen2017pseudo}. 
However, most of the localization methods in \cite{olivier2016lightweight, palacios2017jade, chen2017pseudo} apply only to the indoor  scenarios, and their positioning solutions require supplementary knowledge about the communication environment, such as a floor plan.  
To support the localization of IoT devices in a large-scale open space, an unsupervised approach,  called \emph{channel charting} (CC), was  proposed in \cite{studer2018channel,geng2020multipoint,deng2018multipoint}  to find the relative positions of the served UEs.  
Given the CSI data,  a channel chart captures the propagation features from the radio signals, so as to identify the spatial geometry in the cellular network.  
However, the unsupervised nature of CC  means that an absolute position value cannot be found.  
Meanwhile, all of the prior works in  \cite{studer2018channel,geng2020multipoint,deng2018multipoint} are restricted to multiple-input-single-output transmissions. Thus, these methods cannot be easily adopted for large-scale MIMO communications.

In order to accurately predict location based on MIMO CSI, the prior works in \cite{zhou2017grassma,pan2011tracking}, and \cite{chidlovskii2019semi} improved the fingerprinting approach by adopting a semi-supervised framework, which exploits the network environment using both labeled and unlabeled CSI data.   
The authors in \cite{zhou2017grassma} studied the CSI similarity between different UEs,  and employed the similarity value as the weight of associated location  to determine the UE's position.  
However, this device-centric  method yields a mixed-sized neighboring matrix, which cannot easily be extended to support more UEs as the network size increases.  
To address the flexibility issue,   device-free semi-supervised methods were studied in \cite{pan2011tracking} and \cite{chidlovskii2019semi} . However, both \cite{pan2011tracking} and \cite{chidlovskii2019semi} applied two individual models for unsupervised offline training and supervised online positioning, respectively,  
and this two-model design  degrades the robustness of the proposed localization method to the local training errors.  
Therefore, an integrated model that supports a flexible network size is lacking in the existing literature for  semi-supervised localization.

\subsection{Contributions}
The main contribution of this paper is a novel framework that can perform integrated, parameterized semi-supervised localization in a mmWave  network. 
First, an effective measurement approach is developed to collect multipath CSI allowing each base station (BS) to learn the mmWave network environment.  
Next, an autoencoder model is proposed to estimate the position of an IoT equipment based on its CSI, and an iterative two-stage  algorithm is designed to train the autoencoder  in a semi-supervised manner.  
In the first stage, a channel chart is constructed, using  unlabeled CSI samples, to capture the radio-geometry map of the mmWave system.  
Next, the unsupervised CC model is extended to a semi-supervised framework, where the autoencoder is divided into two components: an encoder and a decoder, and each component is  optimized individually, using the labeled CSI dataset with  associated location information, to further improve  positioning accuracy.  
Simulation results show that the proposed semi-supervised algorithm can capture a better radio-geometry map than a conventional unsupervised CC scheme, and the CC-aided localization approach yields a higher positioning accuracy, compared with existing supervised  methods.      
To the best of our knowledge, this is the first paper that proposes a CC-aided localization framework, using a standalone semi-supervised autoencoder.  

The rest of this paper is organized as follows. 
Section II presents the communication model. 
The CC-aided localization framework  is presented in Section III. 
Simulation results are shown in Section IV. Conclusions are drawn in Section V.

\section{System Model and Data Collection} 

Consider a cellular network, in which a set $\mathcal{B}$  of $B$ BSs provide mmWave communication service to a set $\mathcal{E}$ of UEs.  
We assume  each BS and each UE to be equipped with a uniform linear array (ULA) consisting of $M$ and $N$ antenna elements, respectively, and the adjacent element spacing of the ULA is $d = \lambda/2$, where $\lambda$ is the  carrier wavelength.  
To enable MIMO communications,  pilot training is required to align the beam direction between each BS and its served UE. 
Thus, the multipath channel information of each UE  in  $\mathcal{E}$ will be measured at each BS during the beam training stage.

Let  $\mathcal{E}_l$ and $\mathcal{E}_u $ be two subsets of $\mathcal{E}$, where  $\mathcal{E}_l \cap \mathcal{E}_u = \emptyset$ and $\mathcal{E}_l \cup \mathcal{E}_u = \mathcal{E}$.  
We assume that each UE in $\mathcal{E}_l$ is equipped with a GPS module, and  its location  $\boldsymbol{y}$ is available to the BSs.  
However, for UEs in $\mathcal{E}_u$, their positions are completely unknown,  
thus, each BS only has their wireless channel information.  
Now, consider a new IoT device whose position is unknown, but it requires a location-based service. In order to guarantee an efficient wireless service, each BS aims to measure the multipath parameters towards the IoT device, and then, jointly estimate its location using mmWave CSI.

\subsection{Channel Model}

For a MIMO channel between each BS and wireless equipment,  we consider $L$ different paths.  
Due to the low-scattering propagation feature of mmWave frequencies, the multipath channel is spatially sparse, i.e., $L < MN$.  
The  MIMO channel $\boldsymbol{H} \in \mathbb{C}^{M\times N}$ from a wireless device to a typical  BS will be given by  \cite{abu2018error} \vspace{-0.3cm}  
\begin{equation}
	\boldsymbol{H}(t) =  \sum_{l=1}^L   \beta_l \boldsymbol{a}_r(\phi_l) [ \boldsymbol{a}_t (\psi_l)]^H \delta(t-\tau_l) ,  \vspace{-0.1cm}
\end{equation}
where $\beta_l \in \mathbb{C}$ is the channel gain, $\tau_l$ is ToA of path $l$, 
$ \boldsymbol{a}_r (\phi_l) = [1,  e^{-j\frac{2\pi d}{\lambda} \sin \phi_l }, $ $ \cdots, e^{-j(M-1)\frac{2\pi d}{\lambda}\sin \phi_l }]^T  \in \mathbb{C}^{M\times 1}$ is the  steering vector of the BS with AoA  $\phi_l$, and 
$ \boldsymbol{a}_t (\psi_l) = [1,  e^{-j\frac{2\pi d}{\lambda} \sin \psi_l }, $ $ \cdots,  e^{-j(N-1)\frac{2\pi d}{\lambda}\sin \psi_l }]^T  \in \mathbb{C}^{N\times 1}$ is the transmit steering vector with angle-of-departure (AoD)  $\psi_l$.  

%The single-path channel model $\boldsymbol{H}_l$ for  path $l$ is
%\begin{equation} 
%	\boldsymbol{H}_l =  \beta_l \boldsymbol{a}_r(\phi_l) [ \boldsymbol{a}_t (\psi_l)]^H,  
%\end{equation}
%where $\beta_l \in \mathbb{C}$ is the channel gain, $ \boldsymbol{a}_t (\psi_l) = [1,  e^{-j\frac{2\pi d}{\lambda} \sin \psi_l }, $ $ \cdots,  e^{-j(N-1)\frac{2\pi d}{\lambda}\sin \psi_l }]^T  \in \mathbb{C}^{N\times 1}$ is the transmit steering vector  at the ?? with angle-of-departure (AoD)  $\psi_l$, and 
%$ \boldsymbol{a}_r (\phi_l) = [1,  e^{-j\frac{2\pi d}{\lambda} \sin \phi_l }, $ $ \cdots, e^{-j(M-1)\frac{2\pi d}{\lambda}\sin \phi_l }]^T  \in \mathbb{C}^{M\times 1}$ is the  steering vector of the BS with AoA  $\phi_l$.   

For the BS-UE channel with $L$ distinct paths, each link $l$ has distinct parameters for the  AoD $\psi_l$, AoA $\phi_l$, ToA $\tau_l$, and  channel gain $\beta_l$, which characterize the mmWave communication environment. % and supports CSI-based localization.  
In order to obtain channel parameters  $[ \boldsymbol{\psi}, \boldsymbol{\phi}, \boldsymbol{\tau}, \boldsymbol{\beta} ]$ for all $L$ paths, a transmission model is needed to measure the multipath MIMO channel.

\subsection{Transmission Model and Channel Parameter Estimation}

Given no prior knowledge of location information,  
in order to learn the communication environment and align the beam direction with a mmWave BS,  each UE will apply  a predetermined codebook, consisting of $K$ columns of codewords $\boldsymbol{W} \in \mathbb{C}^{N\times K}$, to explore all possible beam directions \cite{song2018advanced}. 
Each codeword  $\boldsymbol{w}_k$  is a unit-norm  beamforming vector, given by  % \cite{zhou2012efficient}
$	\boldsymbol{w}_k = \frac{1}{\sqrt{N}} \left[1, e^{-j\pi(\frac{2k}{K}-1)}, \cdots, e^{-j\pi(\frac{2k}{K}-1)(N-1)}\right]^T$.  
We assume that each UE has perfect knowledge of its transmit radiation pattern, thus, the transmit beam direction, which is essentially the AoD ${\psi} \in [0,2\pi)$,  can be uniquely determined, based on the beamforming vector $\boldsymbol{w}_k$. 
Given that the codebook $\boldsymbol{W}$ is predetermined, the AoD information for each transmission is available at the BSs \cite{zhang2021dcgan}.

During  beam training, each UE transmits a unit-power pilot signal  $\boldsymbol{s}(t) = [s_1(t), s_2(t), \cdots, s_K(t)]^T$. 
Then, given the directional beamforming matrix  $\boldsymbol{W}$, the transmitted pilot signal from the UE will be $ \sqrt{\rho} \boldsymbol{W} \boldsymbol{s}(t)$, where $\rho$ is the transmission power, and Tr$(\boldsymbol{W}^H\boldsymbol{W}) = 1$.   
Thus, the received  signal observed at the input of the receive beamformer at the BS will be  \vspace{-0.1cm}
\begin{equation}\label{receivedPilot}
	\boldsymbol{r}(t) = \sum_{l=1}^L \sqrt{\rho} \boldsymbol{H}_l \boldsymbol{W} \boldsymbol{s}(t-\tau_l) + \boldsymbol{n}(t),  \vspace{-0.1cm}
\end{equation}  
where $\boldsymbol{H}_l =  \beta_l \boldsymbol{a}_r(\phi_l) [ \boldsymbol{a}_t (\psi_l)]^H$, and $\boldsymbol{n}(t) \sim \mathcal{CN}(\boldsymbol{0}, \sigma^2_{\text{BS} } \boldsymbol{I}_M )$ is the receiver's noise at the BS.

To estimate the AoA value,  each BS needs to steer the receiving beam  and measure the received signal powers in each possible direction \cite{wielandt2017indoor}. 
Indeed, each BS can have its own codebook that determines a set ${\Phi}$ of antenna steering directions.  
To electronically steer the beam  towards each direction ${\phi} \in  {\Phi}$, the BS can linearly combine the received signals at each antenna element with a minimum-variance distortionless response  vector $\boldsymbol{q}({\phi}) = \frac{\boldsymbol{R}^{-1} \boldsymbol{a}_r({\phi})}{\boldsymbol{a}_r({\phi})^H  \boldsymbol{R}^{-1} \boldsymbol{a}_r({\phi})}$, 
where $\boldsymbol{R} = \mathbb{E}\{ \boldsymbol{r}(t) \boldsymbol{r}(t)^H\}$ is the covariance matrix of the received signals.  
Then, the corresponding received power $P({\phi})$ at  steering direction ${\phi}$ can be calculated via $P({\phi}) = \boldsymbol{q}({\phi})^H \boldsymbol{R} \boldsymbol{q}({\phi})$. 
Consequently, the steering directions that yield significant received power will be considered as the estimated AoAs.  
The estimated AoA can provide important information of the signal source and the communication environment. 
In the line-of-sight (LoS) case, the AoA will reveal the direction of the UE's location with respect to the BS, and in the non-line-of-sight (NLoS) scenario, the AoA captures the direction of signal reflection and scattering.

Next,  the value of the channel gain $\beta_l$ for each path $l$   can be calculated based on the received pilot signal and the estimated AoA-AoD information.  
Let $\otimes$ be the Kronecker product, $\circ$ be the  Khatri-Rao product, and vec$(\cdot)$ be the vectorization of a matrix. During the $k$-th training time, with the predetermined beamforming vector $\boldsymbol{w}_k$  and combining vector $\boldsymbol{q}_k$, the received pilot  signal can be given by \cite{han2016two} 
\begin{equation}
	\begin{aligned}
		v_k = \boldsymbol{q}_k^H \boldsymbol{r}_k &=   \sqrt{\rho} \boldsymbol{q}_k^H \boldsymbol{H} \boldsymbol{w}_k {s}_k  + \boldsymbol{q}_k^H\boldsymbol{n}_k \\
		&=   \sqrt{\rho} s_k (\boldsymbol{w}_k^T \otimes \boldsymbol{q}^H_k) \text{vec}(\boldsymbol{H})    + \boldsymbol{q}_k^H \boldsymbol{n}_k \\
		&=  \sqrt{\rho} s_k (\boldsymbol{w}_k^T \otimes \boldsymbol{q}^H_k) (\boldsymbol{A}_t^{\ast} \circ \boldsymbol{A}_r) \boldsymbol{\beta}  + \boldsymbol{q}_k^H \boldsymbol{n}_k, 
	\end{aligned}
\end{equation}
where $\boldsymbol{\beta} = [\beta_1, \cdots, \beta_L]$, $\boldsymbol{A}_t = [\boldsymbol{a}_t(\psi_1), \cdots, \boldsymbol{a}_t(\psi_L)]$, and $\boldsymbol{A}_r = [\boldsymbol{a}_r(\phi_1), \cdots,  \boldsymbol{a}_t(\phi_L)]$.  
Based on the estimated AoD $\psi_l$  and AoA $\phi_l$, the values of $\boldsymbol{A}_t$ and $\boldsymbol{A}_r$ are known. %  in Section \ref{aod} and \ref{aoa}
Then, by stacking $K$ observations, we have  
\begin{equation}
	\boldsymbol{v} = \sqrt{\rho} \boldsymbol{s} (\boldsymbol{W} \circ \boldsymbol{Q}^{\ast})^T (\boldsymbol{A}_t^{\ast} \circ \boldsymbol{A}_r) \boldsymbol{\beta}  + \bar{\boldsymbol{n}},  
\end{equation} 
where $\boldsymbol{Q} = [\boldsymbol{q}_1, \cdots, \boldsymbol{q}_K]$, and $\bar{\boldsymbol{n}} = [\boldsymbol{q}_1^H \boldsymbol{n}_1, \cdots, \boldsymbol{q}_K^H \boldsymbol{n}_K ] $. 
Thus, the channel gain can be estimated via $\tilde{\boldsymbol{\beta}} = (\boldsymbol{F}^T \boldsymbol{F})^{-1} \boldsymbol{F}^T \boldsymbol{v}$, where $\boldsymbol{F} =  \sqrt{\rho} \boldsymbol{s} (\boldsymbol{W} \circ \boldsymbol{Q}^{\ast})^T (\boldsymbol{A}_t^{\ast} \circ \boldsymbol{A}_r) \in \mathbb{C}^{K\times L}$ and $K\ge L$.   
The norm of the channel gain represents the received signal strength, which indicates the distance-based path loss, possible power loss due to reflection, and mmWave fading parameters, while  the phase of the channel gain can reveal possible phase shifts caused by scattering.  
Thus, the complex value of $\boldsymbol{\beta}$ provides more information about the network environment.

Furthermore, we assume that a reference clock is available for time synchronization in our system.    
Since the signal travel time is a result of the propagation distance divided by the speed of light, it provides an accurate estimation of the path length.   
Given multiple BSs, the intersection of multiple circles that are centered at each BS with a measured propagation distance can determine the  spatial area where the wireless device is possibly located. % \cite{oumar2012comparison}. 
Consequently, after the beam training stage, each BS can collect a state vector of the considered IoT device, i.e., the estimated channel parameters $\boldsymbol{u} = [\boldsymbol{\psi}, \boldsymbol{\phi}, \boldsymbol{\tau}, |{\boldsymbol{\beta}}|, \angle {\boldsymbol{\beta}} ] \in \mathbb{C}^{1\times 5L}$.

\subsection{Problem Statement}

Given the measured channel information, in order to map the multipath CSI into a 3D location in the communication network, each  BS $b \in \mathcal{B}$ applies two CSI datasets, where  
the labeled dataset $\mathcal{L}_b = \{ \boldsymbol{u}_{n,b} , \boldsymbol{y}_{n}\}_{n \in \mathcal{E}_l }$  has both the multipath data $\boldsymbol{u}_{n,b}$ and location information $\boldsymbol{y}_n$ of each labeled UE $n \in\mathcal{E}_l$,   
and the unlabeled dataset $\mathcal{U}_b = \{ \boldsymbol{u}_{n,b} \}_{n \in \mathcal{E}_u}$ only contains the channel parameters.   
Therefore, given both CSI datasets $\mathcal{L}= \{ \mathcal{L}_b \}_{b\in \mathcal{B}}$ and $\mathcal{U} = \{ \mathcal{U}_b \}_{b\in \mathcal{B}}$ for all BSs, the objective of  CSI-based localization is to find the position of the considered IoT equipment, based on its multipath information $\boldsymbol{u}$ at each BS.

\section{Channel Charting Aided Localization}\label{SecLearning}

The use of both labeled and unlabeled datasets  to improve the positioning accuracy is our main objective.  
To this end, we first apply the unlabeled  data $\mathcal{U}$ to build a channel chart  that maps the high-dimensional CSI into a low-dimensional chart component, while preserving the local geometry feature of the real  location for each UE \cite{studer2018channel}. %spatial
The CC captures  the spatial relations  of different UEs by identifying the similarity of their CSI samples in an unsupervised manner, such that the neighboring UEs in the spatial location remain close in the channel chart. 
Next,  this unsupervised model is extended to a semi-supervised learning framework, by using the associated location information in the labeled dataset $\mathcal{L}$,  to estimate the absolute value of the IoT device's location.   
 
\begin{figure}[!t]
	\begin{center}  
		\includegraphics[width=8cm]{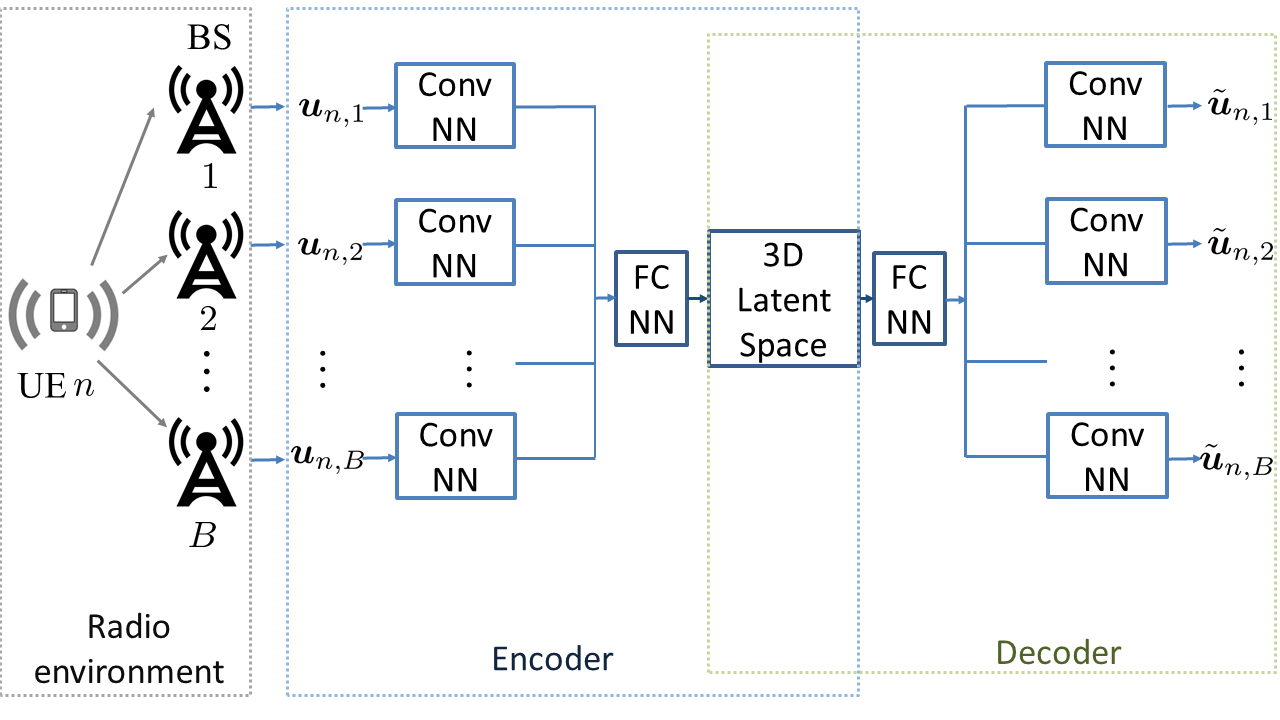} 
		\caption{\label{learningFramework}\small An autoencoder-based CC framework for semi-supervised localization of wireless devices. }  
	\end{center}\vspace{-0.7 cm}  
\end{figure}
  
\vspace{-0.05cm}
\subsection{Channel Charting} \label{cc}

Fig. \ref{learningFramework} presents the framework of CC mapping, in which each BS $b$ collects multipath information $\boldsymbol{u}_{n,b}$ for each UE $n$, and then, feeds the  sample to a \emph{convolutional autoencoder}. 
The autoencoder applies a convolutional neural network (CNN) based framework for unsupervised learning, which includes an encoder and a decoder. 
The encoder $\mathcal{C}_{\Omega}:$ $\mathbb{R}^{B\times 5L} \rightarrow \mathbb{R}^{3} $ receives a high-dimensional CSI sample as its input, and implements a low-dimensional representation  $\tilde{\boldsymbol{y}}_n = \mathcal{C}_{\Omega}(\boldsymbol{U}_n)$ that captures the essential 3D location information of UE $n$, where $\Omega$ is the   parameter of the encoder, and $\boldsymbol{U}_n = [\boldsymbol{u}_{n,1}^T, \cdots, \boldsymbol{u}_{n,B}^T]^T$.  
Following that, a decoder $\mathcal{D}_{\Theta}:  \mathbb{R}^{3} \rightarrow \mathbb{R}^{B\times 5L} $ with parameter $\Theta$ is applied to reconstruct the CSI $\tilde{\boldsymbol{U}}_{n}$ from the location representation  $\tilde{\boldsymbol{y}}_n$ for UE $n$.  
The purpose of a decoder is to reconstruct CSI parameters that approximates  the input sample, i.e.,  $  \mathcal{D}_{\Theta}(\mathcal{C}_{\Omega}({\boldsymbol{U}}_n )) =\tilde{\boldsymbol{U}}_{n} \approx {\boldsymbol{U}}_{n}  $.  
To achieve this purpose and facilitate the performance evaluation of the autoencoder-based CC, the unlabeled CSI samples in  $\mathcal{U}$ are used for both the input and the training output. %, where  the encoder first reduces a CSI sample to a latent 3D location, and, then, the decoder maps from this estimated 3D position back to CSI. 

As shown in Fig. \ref{learningFramework}, in the encoder part, the CSI sample is first processed by a local CNN  at each BS to capture the local radio-geometry pattern. Then, all extracted information from different sources is fused together through a fully-connected neural network (FCNN)   at a central learner, where the  overall location properties of the considered equipment will be jointly evaluated, and finally  the 3D code of the latent channel chart is extracted.   
Next, the decoder takes the location information in the CC space, learn the radio-geometry features near the considered equipment, and then, reconstruct the multipath parameters observed by BSs given the position of each BS.      
Consequently, the loss function of the overall CC framework given by:  \vspace{-0.05cm}
\begin{equation}
	E(\Omega,\Theta) = \frac{1}{2U}  \sum_{n=1}^{U} \|{\boldsymbol{U}}_{n} - \mathcal{D}_{\Theta}(\mathcal{C}_{\Omega}({\boldsymbol{U}}_{n} )) \|_2^2 + \frac{\beta}{2} \|  \Omega_{\text{FC}}\|_F^2,  \vspace{-0.15cm} 
\end{equation}
where $U$ is the number of unlabeled CSI samples in $\mathcal{U}_b$ for each BS,  $\beta>0$ is the tuning parameter for the squared Frobenius regularizer, and $\Omega_{\text{FC}}$ is the FCNN parameter  in the encoder. 
In the training phase, the Adam optimizer is used to minimize the loss function $E$ over $\Omega$ and $\Theta$.

\subsection{Channel Charting-Aided Location Estimation} \label{sup}

Given the unsupervised CC mapping,  we will extend the  autoencoder model  to support a  semi-supervised learning framework. Thus, labeled CSI samples whose 3D latent  representation is given in $\mathcal{L}$ can be employed to further improve the positioning accuracy of the CC-aided localization approach. 

As mentioned before, the encoder itself can form a CNN-based positioning framework, where the input is the multipath channel sample $\{\boldsymbol{u}_{n,b}\}_{\forall b}$, and the output is the estimated 3D location $\tilde{\boldsymbol{y}}_{n}$ of  UE $n$. 
Meanwhile, the decoder can be considered as a channel  predictor, where given the location ${\boldsymbol{y}}_{n}$ as input, the multipath parameter  $\tilde{\boldsymbol{u}}_{n,b}$ can be generated as output at each BS $b$. 
Therefore, using the labeled dataset $\mathcal{L} $, we can further train the encoder and decoder separately in a supervised manner.  
That is, after the CC training using the unlabeled dataset $\mathcal{U}$, we can divide the autoencoder model into an  encoder and a decoder for UE localization and channel prediction, respectively, and  each component can be optimized further using labeled CSI samples with associated location information.

In the supervised training of the encoder, we aim to estimate the location of the considered device given the multipath CSI, i.e.,  $ \mathcal{C}_{\Omega}(\boldsymbol{U}_n) = \tilde{\boldsymbol{y}}_n  \approx {\boldsymbol{y}}_n $. 
Therefore, the loss function for the encoder-based localization task is 
\begin{equation} 
	E_{\text{e}}(\Omega) = \frac{1}{2J}  \sum_{n=1}^{J} \|{\boldsymbol{y}}_{n} - \mathcal{C}_{\Omega}({\boldsymbol{U}}_{n} ) \|_2^2 + \frac{\beta}{2} \|  \Omega_{\text{FC}}\|_F^2, 
\end{equation} 
where  $J$ is the number of labeled samples in $\mathcal{L}_b$ for each BS. 
Then, in the training stage, parameter $\Omega$  will be optimized to minimize the loss function $E_{\text{e}}$ for the encoder. % 

Meanwhile, the purpose of the decoder is to generate the multipath channel sample at each BS, given the known location of the considered device, i.e., $ \mathcal{D}^{\Theta}(\boldsymbol{y}_n) = \tilde{\boldsymbol{U}}_n  \approx {\boldsymbol{U}}_n $. 
Thus, the approximate error for the decoder's supervised learning  is 
\begin{equation} 
	E_{\text{d}}(\Theta) = \frac{1}{J}   \sum_{n=1}^{J} \|{\boldsymbol{U}}_{n} - \mathcal{D}_{\Theta}({\boldsymbol{y}}_{n} ) \|_2^2 .
\end{equation} 
Then, an optimization process will be applied to minimize $E_{\text{d}}$ over $\Theta$.   
To the best of our knowledge, this is the first work that applies a CC framework to assist semi-supervised localization, using a standalone autoencoder model.  
The  training process for the proposed CC-aided localization approach is summarized in Algorithm \ref{algo222}.

\begin{algorithm}[h] %\small   
	\caption{Channel charting-aided localization } \label{algo222}
	\begin{algorithmic}
		\State \textbf{Initialize} $\Omega$ and $\Theta$ for encoder and decoder, respectively;  \\  %\cite{vincent2008extracting}
		\textbf{Determine} minibatch sizes $B_u$ and $B_l$ for $\mathcal{U}$ and $\mathcal{L}$, respectively, such that $\frac{U}{B_u} = \frac{L}{B_l} = N$; \\		
		\textbf{For} $i$ in $[1, \cdots, N]$, \textbf{do}\\
		\quad \textbf{For} $b$ in $[1, \cdots, B_u]$, \textbf{do}\\
		\quad \quad $\Omega, \Theta \leftarrow \arg\min_{\Omega,\Theta} E$; // unsupervised CC mapping\\
		\quad \textbf{For} $b$ in $[1, \cdots, B_l]$, \textbf{do}\\
		\quad \quad $\Omega \leftarrow \arg\min_{\Omega} E_e$; // supervised encoder training\\
		\quad \textbf{For} $b$ in $[1, \cdots, B_l]$, \textbf{do}\\
		\quad \quad $\Theta \leftarrow \arg\min_{\Theta} E_d$; // supervised decoder training\\ 
		\textbf{Return} $\Omega$ and $\Theta$. 
	\end{algorithmic}
\end{algorithm}

\section{Simulation Results and Analysis}

\begin{figure}[!t]
	\begin{center}
		\vspace{-0.15cm}
		\includegraphics[width=8.9cm]{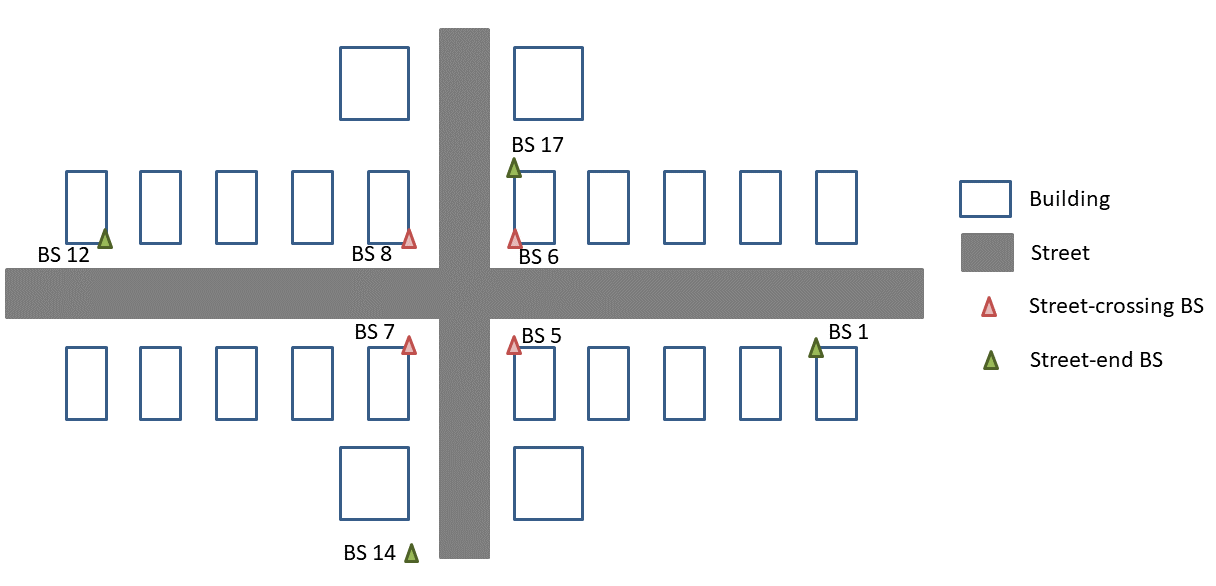}
		\vspace{-0.1cm}
		\caption{\label{simulationsetting}\small A top view of the simulation environment in an outdoor communication scenario, where the  street length is $550$ meters \cite{alkhateeb2019deepmimo}.  }  
	\end{center}\vspace{-0.5cm}  
\end{figure}

In our simulations, we  consider $B=4$ BSs with $M=256$ antennas. Each UE is equipped with $N=64$ antennas.   
To evaluate the performance of the proposed CC-aided localization approach, we use the open-source DeepMIMO dataset in \cite{alkhateeb2019deepmimo} that applies ray-tracing techniques to measure the AoA, AoD, ToA,  phase, and power of channel gain for  $28$ GHz  communications in an outdoor street scenario with up to $L=5$ multipaths.   
We use  $J = 30,000$ labeled CSI samples, and $U =72,000 $ unlabeled data. 
For both datasets,  $85\%$ samples is employed for model training, and the remaining $15\%$ for testing. 
In our simulations, both LoS  and NLoS channel states for mmWave MIMO communications are studied. 
As shown in Fig. \ref{simulationsetting}, to support a LoS-dominant communication setting, we choose  BSs $5$, $6$, $7$, and $8$, which are located near the crossing of two streets, to collect the CSI samples, where a LoS path exists for most of the mmWave links.  
For the NLoS-dominant scenario, we use  BSs $1$, $12$, $14$, and $17$ located at the end of each street to collect the channel parameters, thus most paths in the measured  channels are reflected and NLoS.     

For the  autoencoder model, the encoder has four hidden layers, where the first three layers are convolutional with rectified linear unit (ReLU) activations, each of which is followed by a max-pooling operation, and the last hidden layer is fully-connected with a sigmoid activation. The decoder has a mirror structure of the encoder.

\subsection{Performance of Channel Charting} 
To evaluate the accuracy of CC mapping, we consider two standard metrics, i.e., continuity (CT) and trustworthiness (TW), which are widely used for dimensionality reduction \cite{studer2018channel}.   
Both  CT and TW have a value range within $[0, 1]$, and a larger value indicates a better CC mapping result. 
To define theses two metrics, first,  we introduce $\mathcal{V}^K(n)$ as the set of $K$-nearest neighbors of equipment $n$ in the real physical space, and we define $r_v(n,i)$ as the ranking of device $i$ among the neighbors of $n$ based on their real-world 3D locations. 
In the CC latent space, we define $\mathcal{F}^K(n)$ to be the set of ``false neighbors'' that stay in the $K$-nearest neighbors of $n$ in the latent space, but not of $n$ in the original location space. Meanwhile, $r_f(n,i)$ is the ranking of device $i$ to device $n$ ranked by their latent locations in the channel chart. 
Then, CT is defined to measure on whether neighbors of each equipment in the real physical space are preserved in the latent charting space, i.e.,
\begin{equation*} 
	CT(K) = 1 - \frac{2}{UK(2U-3K-1)}\sum_{n=1}^U \sum_{i \in \mathcal{V}^K_n} [r_f(n,i)-K],   
\end{equation*} 
where a large CT value (close to $1$) means that all of the original neighbors of each UE remain  close to it  after CC mapping, thus, the latent space  preserves the neighboring relationship of wireless UEs.   
Next, TW is defined to measure how well the CC mapping avoids introducing false neighbor relations that were absent in the original space, i.e.,  
\begin{equation*} 
	TW(K)  = 1 - \frac{2}{UK(2U-3K-1)} \sum_{n=1}^U \sum_{j\in \mathcal{F}^K_n}  [r_v(i,j) - K].  
\end{equation*}  
Here, a large TW value (close to $1$) indicates that no false neighbor is introduced after CC mapping.  

\begin{figure}[!t]
	\begin{center}
		\vspace{-0.13cm}
		\includegraphics[width=9cm]{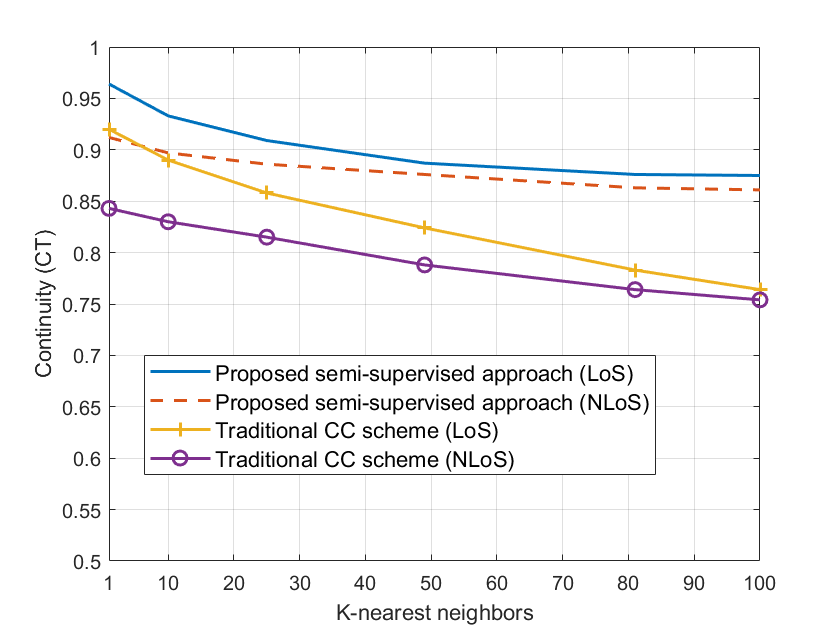}
		\vspace{-0.2cm}
		\caption{\label{ct}\small The value of continuity as the neighbor size  $K$ increases. }  
	\end{center}\vspace{-0.4cm}  
\end{figure}

\begin{figure}[!t]
	\begin{center}
	    \vspace{-0.13cm}
		\includegraphics[width=9cm]{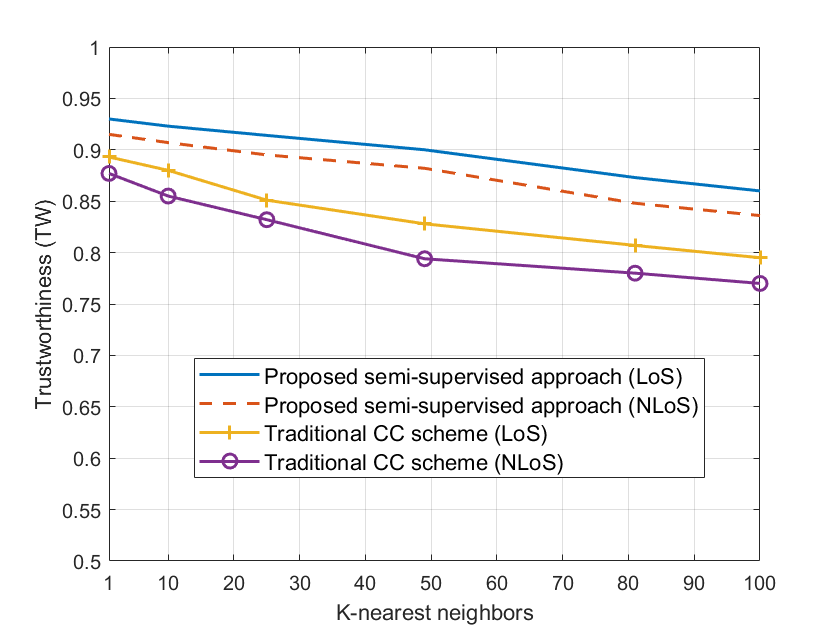}
		\vspace{-0.2cm}
		\caption{\label{tw}\small The value of trustworthiness as the neighbor size  $K$ increases. }  
	\end{center}\vspace{-0.4cm}  
\end{figure}

In Figs. \ref{ct} and \ref{tw}, we compare the values of TW and CT, respectively, as the neighbor size $K$ increases from $1$ to $100$. 
In order to evaluate the performance of our proposed semi-supervised  approach, an unsupervised autoencoder is introduced as a baseline, which employs the same network topology in Fig. \ref{learningFramework}, but  it does not access the labeled data in $\mathcal{U}$ for model training.  
First, as shown in Fig. \ref{ct}, our proposed CC method yields a larger CT value, compared with the baseline  scheme, in both the LoS and NLoS communication scenarios.   
The proposed semi-supervised CC approach maps each device  from the physical radio space to the 3D latent space while accurately preserving the neighboring relationship, such that most of the devices still have their original $K$-nearest neighbors as their closest devices in the CC latent space. 
However, as the value of $K$ increases, the values of CT for both approaches decrease. 
As more neighbors are counted and compared between the original geometry space and the latent channel chart, it becomes more likely for an original neighbor to be mistakenly ranked lower after the CC mapping. Thus, the value of CT becomes smaller, given a larger neighbor size  $K$.  

Compared with the baseline CC approach, the CT value of the proposed semi-supervised method decreases more slowly, for two reasons. 
First, the baseline CC only uses the unlabeled dataset $\mathcal{U}$ to train the autoencoder, while our proposed approach can access two datasets $\mathcal{U}$  and $\mathcal{L}$ for more CSI samples. 
Given a larger size of training data, our proposed CC framework can capture an accurate pattern of CSI values associated with the location of each device.  
Thus, the spatial relationship can be better represented in the latent space, compared with the traditional CC algorithm. 
Next, the location information in the labeled dataset provides one more metric when determining the CSI similarity between multiple UEs. Therefore, the performance of semi-supervised learning is better and more robust, compared with the unsupervised CC model.  
For  similar reasons, our proposed method also outperforms the baseline CC scheme, in the evaluation of TW value, as shown in Fig. \ref{tw}. A larger TW value indicates that the proposed CC approach   introduce less false neighbors to each UE in the latent CC space. 
In  summary, compared with the traditional CC scheme that only applied unlabeled data for unsupervised learning, our proposed semi-supervised framework can improve the accuracy   of CC mapping, using the location information in  labeled CSI samples.

\subsection{Performance of Localization }

\begin{figure}[!t]
	\begin{center}
		\vspace{-0.13cm}
		\includegraphics[width=9cm]{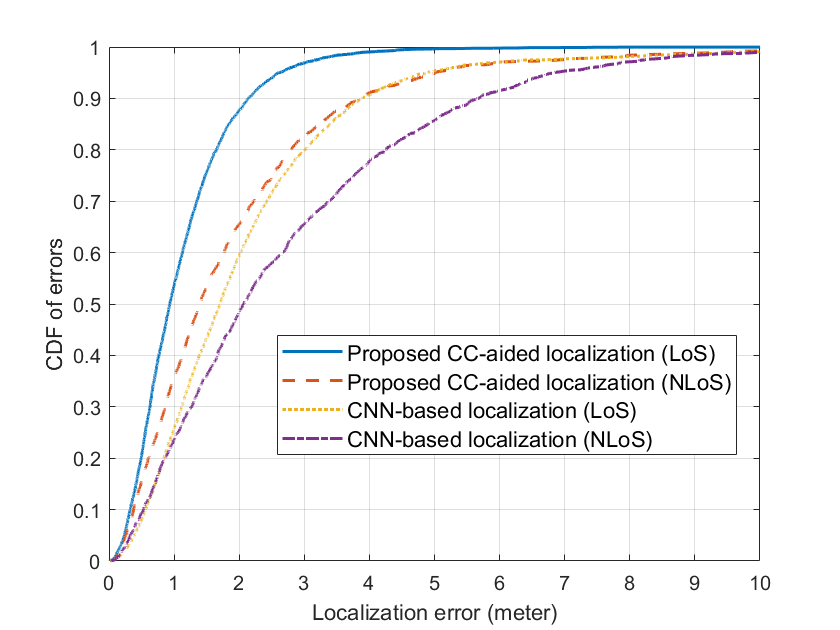}
		\vspace{-0.2cm}
		\caption{\label{cdfLocError}\small The CDF of localization errors. }  
	\end{center}\vspace{-0.4cm}  
\end{figure}

In Fig. \ref{cdfLocError}, we evaluate the positioning accuracy of the proposed CC-aided localization method.  
For performance comparison, we introduce a baseline scheme that employs a CNN framework to predict the IoT device's location based on CSI. The baseline model has the same structure as the encoder in Section \ref{sup}, which only applies the labeled dataset $\mathcal{L}$ for model training, but it does not know any information about the channel chart.  %. However, the baseline scheme
We use the norm-$2$ distance between the predicted location and real location $\|\boldsymbol{y} - \tilde{\boldsymbol{y}} \|_2$  as the performance metric.  

Fig. \ref{cdfLocError}  shows that   our proposed CC-aided method outperforms the baseline scheme.  
First, in the LoS communication scenario, to achieve a localization error that is smaller than $2$ meters, our proposed positioning approach shows an empirical probability of over $85\%$, while the baseline scheme only reaches $60\%$. 
When the mmWave paths are dominated by NLoS links, the positioning accuracy of both approaches decreases, because an NLoS path cannot directly provide the angular information of the IoT device's  orientation. 
However, the CC mapping can capture the neighboring relationship between different wireless devices, such that the BSs can jointly learn the propagation features of surrounding environment based on  multi-source CSI.   
Thus, the CC-aided localization scheme can obtain the reflection information  of mmWave links to adjust the NLoS  parameters, and then, infer the IoT's position.    
Due to the aid of CC, our proposed semi-supervised localization scheme yields a smaller positioning error in both the LoS and NLoS communication scenarios, compared with a conventional supervised learning baseline.

%\subsection{Performance of User Tracing }

\section{Conclusion}
 
In this paper, we have proposed a novel  framework for semi-supervised localization in a mmWave network.
In particular, we have developed a convolutional autoencoder model to enable a CC-aided positioning using both labeled and unlabeled CSI dataset.  
In order to capture the radio-geometry map of the mmWave system, we have built a channel chart in an unsupervised manner. 
Next, we have divided the autoencoder model into two components, each of which was  trained individually, using the labeled CSI dataset with  associated location information,  to further improve the positioning accuracy.  
Simulation results show that the proposed semi-supervised learning approach captures a better radio-geometry map than an unsupervised CC scheme. Meanwhile, our proposed CC-aided localization method yields a higher positioning accuracy, compared with a conventional supervised learning scheme.

\bibliographystyle{IEEEtran}
\bibliography{references}

\end{document}